\documentclass[pra,twocolumn,superscriptaddress,longbibliography]{revtex4-1}
\usepackage{amsmath,amssymb,graphicx,bm,braket,color}
\usepackage{hyperref}

\def\br{{\bf r}}
\def\bk{{\bf k}}
\def\bq{{\bf q}}
\def\bpsi{{\bar\psi}}
\def\bchi{{\bar\chi}}
\def\hz{{\bf \hat z}}

\begin{document}
\title{Nodal-line pairing with 1D-3D coupled Fermi surfaces: a model
  motivated by Cr-based superconductors}

\author{Gideon Wachtel} 
\affiliation{Department of Physics, University of Toronto, Toronto,
  Ontario M5S 1A7, Canada}

\author{Yong Baek Kim} 
\affiliation{Department of Physics, University of Toronto, Toronto,
  Ontario M5S 1A7, Canada}
\affiliation{Canadian Institute for Advanced Research/Quantum
  Materials Program, Toronto, Ontario MSG 1Z8, Canada}
\affiliation{School of Physics, Korea Institute for Advanced Study,
  Seoul 130-722, Korea}

\begin{abstract}
  Motivated by the recent discovery of a new family of Chromium based
  superconductors, we consider a two-band model, where a band of
  electrons dispersing only in one direction interacts with a band of
  electrons dispersing in all three directions. Strong $2k_f$ density
  fluctuations in the one-dimensional band induces attractive
  interactions between the three-dimensional electrons, which, in turn
  makes the system superconducting.  Solving the associated Eliashberg
  equations, we obtain a gap function which is peaked at the ``poles''
  of the three-dimensional Fermi sphere, and decreases towards the
  ``equator''. When strong enough local repulsion is included, the gap
  actually changes sign around the ``equator'' and nodal rings are
  formed. These nodal rings manifest themselves in several
  experimentally observable quantities, some of which resemble
  unconventional observations in the newly discovered superconductors
  which motivated this work.
\end{abstract}

\maketitle

\section{Introduction}

Recent experiments\cite{bao_superconductivity_2015,
  tang_unconventional_2015, tang_superconductivity_2015} have
discovered superconductivity in a new family of Chromium based
compounds, K$_2$Cr$_3$As$_3$, Rb$_2$Cr$_3$As$_3$, and
Cs$_2$Cr$_3$As$_3$, which consist of well separated Cr$_3$As$_3$
chains. As sometimes expected for one-dimensional systems, enhanced
heat capacity\cite{bao_superconductivity_2015, kong_anisotropic_2015}
and unusual nuclear magnetic resonance (NMR)
measurements\cite{zhi_nmr_2015} in the normal state indicate that
electron-electron interactions play an important role.  In the
superconducting phase, experiments show signatures of an
unconventional pairing mechanism, with nodes in the gap
function. Among these are the absence of a Hebel-Slichter peak in NMR
measurements\cite{zhi_nmr_2015}, the linear decrease of the superfluid
stiffness as seen in muon spin relaxation
($\mu$SR)\cite{adroja_superconducting_2015} and penetration
depth\cite{pang_evidence_2015} experiments, and $\sqrt{H}$ increase in
the specific heat\cite{tang_unconventional_2015} under an applied
magnetic field $H$.

The electronic band structures of K$_2$Cr$_3$As$_3$ and
Rb$_2$Cr$_3$As$_3$ have been calculated\cite{xian-xin_magnetism_2015,
  jiang_electronic_2015}, using Density Functional Theory (DFT), which
finds a single three-dimensional (3D) Fermi-surface, and several
quasi-one-dimensional (Q1D) Fermi surfaces, with electronic dispersion
primarily along the Cr$_3$As$_3$ chains. Some of the Q1D Fermi
surfaces nearly touch the 3D Fermi surface.  We emphasize here the
uniqueness of this band structure, as compared with other
quasi-one-dimensional superconductors.  While the electronic band
structure is highly anisotropic in other Q1D
superconductors\cite{armici_new_1980, jerome_organic_1982}, in the
Chromium based superconductors, anisotropic Q1D bands coexist, and
interact, with a conventional, 3D band. Motivated by this observation,
we ask, what kind of superconductivity can arise from coupling between
electrons in Q1D and 3D bands.

\begin{figure}[h]
  \centering
  \includegraphics[width=\linewidth]{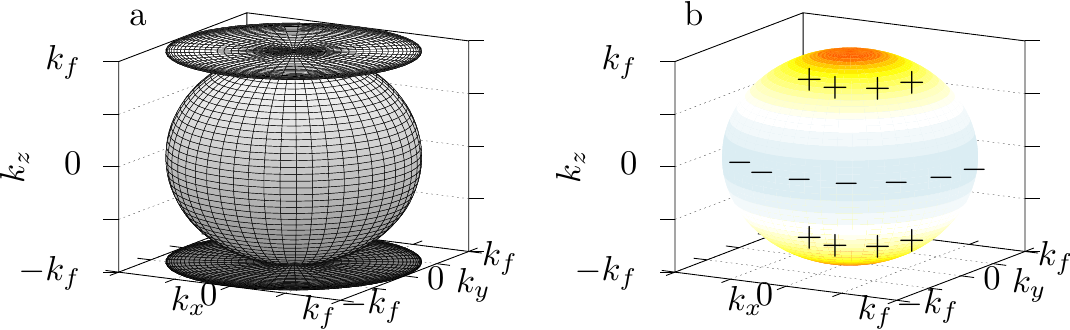}
  \caption{ Left: Fermi Surface geometry of the model,
    Eq. (\ref{eq:S0}). The 1D Fermi sheets touch the 3D Fermi sphere
    at its ``north and south poles''. A $2k_f$ momentum transfer
    between the two 1D band sheets, is also an allowed momentum
    transfer between these points. Right: Gap function with nodal
    rings, plotted on the 3D Fermi shpere. Yellow-red regions
    indicate $\Delta(z)>0$, while blue indicates $\Delta(z)<0$. The
    gap nodes are depicted by the two white rings around the Fermi
    surface in the upper and lower hemispheres. }
  \label{fig:FS}
\end{figure}

To answer this question, we introduce a simplified model, consisting
of a single one-dimensional (1D) band, with dispersion in only one
direction, and a 3D band which disperses isotropically in all
directions. The 1D Fermi-surface sheets touch the Fermi sphere at its
``north and south poles'', as depicted in figure
\ref{fig:FS}a. One-dimensional electronic systems exhibit enhanced
density fluctuations with a wave vector $2k_f$ where $k_f$ is the
Fermi momentum. A $2k_f$ wave vector also connects the two ``poles''
of the 3D Fermi sphere, allowing the 1D fluctuations to induce a
strong attractive force between the opposite points. This effective
attractive interaction is strongly dependent on momentum and energy,
and can therefore give rise to pairing gap functions of an anisotropic
$d_{z^2}$-wave nature, in addition to the more conventional uniform
$s$-wave gap functions.  Higher order functional forms are possible as
well. When the local repulsion within the 3D band is strong enough to
suppress $s$-wave superconductivity, a $d_{z^2}$-wave like gap
function is expected to describe the leading supercoducting
instability of the model. We show that indeed this is the case by
numerically solving the Eliashberg equations for the effective
interaction.

A gap function of a $d_{z^2}$-wave functional form is characterized by
a change in sign as one moves away from the ``poles'' on the Fermi
sphere, towards the ``equator'', see figure \ref{fig:FS}b. The points
where the gap changes sign form two gapless nodal rings in the upper
and lower Fermi hemispheres. The presence of such nodes in the gap
function, can manifest itself in various experimental signatures. We
use the gap functions we obtained from solving the Eliashberg
equations to show that our model is indeed expected to show
unconventional behavior in NMR and superfluid stiffness measurements,
similar to those observed in the Chromium based superconductors.
Since our results rely on generic density-density interactions, the
$d_{z^2}$-wave nodal ring pairing would be a viable candidate for the
newly discovered superconducting phase in these materials.

\section{1D-3D coupled Fermi surfaces}

Consider a three-dimensional system with a band of spin half fermions
(denoted by $\chi$) dispersing only in one direction ($k_z$) and an
additional band of spin half fermions ($\psi$) dispersing
isotropically in all  directions. The zero temperature imaginary time
action of such a system is given by
\begin{eqnarray}
  \label{eq:S0}
  S_0 & = & \sum_\sigma\int_{\bk,\omega}\bchi_\sigma(\bk,\omega)
  \left(-i\omega+\frac{1}{2m} (k_z^2-k_f^2)\right)\chi_\sigma(\bk,\omega) \nonumber \\
  & & \!+\! \sum_\sigma\int_{\bk,\omega}\bpsi_\sigma\!(\bk,\omega)\left(\!-i\omega
    +\frac{1}{2m}\!
    (|\bk|^2\!-\!k_f^2) \! \right)\!\psi_\sigma(\bk,\omega), \nonumber \\ & & 
\end{eqnarray}
where $\sigma=\uparrow,\downarrow$. $k_f$ and $m$ are, respectively,
the Fermi momentum and fermion mass, which, we take to be identical
for both bands.  We further consider local density-density repulsion
between the fermions, as given by
\begin{equation}
  \label{eq:SI}
  S_I  =  S_I^{\rm 1D}
  + U\int_{\br,\tau}\bpsi_\uparrow\psi_\uparrow\bpsi_\downarrow
    \psi_\downarrow 
    + V\sum_{\sigma,\sigma'}\int_{\br,\tau}
    \bchi_\sigma\chi_\sigma \bpsi_{\sigma'}\psi_{\sigma'}. 
\end{equation}
Here, $U$ sets the repulsion within the 3D band, $V$ the repulsion
between the two bands, and $S_I^{\rm 1D}$ allows for a general form of
interactions within the 1D band.

In this paper we study the effect of these interactions on the 3D
fermions, and therefore trace out the 1D fermions (we discuss
the feedback of the 3D fermions onto the 1D band in Appendix
\ref{sec:app}). The leading effect comes from the effective
interaction between 3D fermions as obtained by second order
perturbation in $V$,
\begin{eqnarray}
  \label{eq:Seff}
  S_{\rm eff} & = & -\frac{V^2}{2}\sum_{\sigma,\sigma'}\int_{\bk\bk'\bq,\omega\omega'\Omega}
  \bpsi_\sigma(\bk+\bq,\omega+\Omega)\psi_\sigma(\bk,\omega) 
  \nonumber \\ & & \qquad 
  C(\bq,\Omega) 
  \bpsi_{\sigma'}(\bk'-\bq,\omega'-\Omega) \psi_{\sigma'}(\bk',\omega') .
\end{eqnarray}
$C(\bq,\Omega)$ is the Fourier transform of the density-density
correlation function of the 1D band,
\begin{equation}
  \label{eq:Cdef}
  C(\br,\tau)\! = \! \braket{\bchi(\br,\tau)\chi(\br,\tau)
    \bchi(0,0)\chi(0,0)}_{V=0} \!
  -\braket{\bchi\chi}^2_{V=0}.
\end{equation}
In the absence of interactions, $S_I^1=0$, it is given by
\begin{equation}
  \label{eq:Cq}
  C_0(\bq,\Omega)=\frac{Am}{2\pi q_z}\ln\frac{\Omega^2+(2k_fq_z+q_z^2)^2/4m^2}
  {\Omega^2+(2k_fq_z-q_z^2)^2/4m^2},
\end{equation}
where $A$ is proportional to the Brillouin zone cross section,
perpendicular to the $k_z$ axis. $C_0(\bq,\Omega)$ has a logarithmic
divergence for $q_z=\pm2k_f$. When interactions within the 1D band are
turned on, the $\chi$ fermions form a Luttinger liquid, with power law
correlations.  Had the 1D band been spinless, the $2k_f$ divergence
would have had the form,
\begin{equation}
  \label{eq:CLL}
  C(\bq,\Omega) \simeq \left[\Omega^2 + v^2(q_z \pm 2k_f)^2\right]^{K-1},
\end{equation}
where the effective Luttinger parameter $K$, as well as the velocity
$v$ associated with 1D charge fluctuations, are determined by the
interaction terms in $S_I^{\rm 1D}$. However, in the spinfull case
there are two velocities, $v_c$ and $v_\sigma$, for charge and spin
excitations, respectively. Generally, $v_c\ne v_\sigma$, and the exact
form of the correlation function is unknown, due to a lack of Lorentz
invariance\cite{orgad_spectral_2001, iucci_fourier_2007}.
Nevertheless, we shall assume $v_c=v_\sigma=v$ for
simplicity. Qualitatively, our results depend only on the existence of
enhanced $2k_f$ interactions, while their exact form is less
important. Starting from the non-interacting case, $K=1$, the
parameter $K$ is still expected to decrease for stronger repulsion
within the 1D band. Thus, the sharpness of the interaction is tuned by
the strength of these repulsive forces.

\section{The Eliashberg equations}

The main purpose of this paper is to study the superconducting phase
which can emerge from the effective interaction,
Eq. (\ref{eq:Seff}). To do so, one must solve the associated
Eliashberg equations, since, this interaction depends strongly on
frequency. The task is somewhat simplified if one approximates the
interaction by its projection onto the Fermi surface, i.e., by mapping
momenta $\bk$ onto their polar angles $\theta$, measured with respect
to the positive $k_z$ axis. Thus, the coupled \emph{finite
  temperature} Eliashberg equations take the form
\begin{eqnarray}
  \label{eq:Elfull1}
  \Sigma(z,i\omega_n) & = & g\, T\sum_{i\omega_m}\int_{-1}^1 dz'\,
  \lambda(z-z',\omega_n-\omega_m)\nonumber \\ & & 
  \frac{\omega_m+\Sigma(z',i\omega_m)}
  {\sqrt{(\omega_m+\Sigma(z',i\omega_m))^2+\Phi^2(z',i\omega_m)}} \\ 
  \Phi(z,i\omega_n) & = & g\,T\sum_{i\omega_m}\int_{-1}^1 dz'\,
  \lambda(z-z',\omega_n-\omega_m) \nonumber \\ & & \frac{\Phi(z',i\omega_m)}
  {\sqrt{(\omega_m+\Sigma(z',i\omega_m))^2+\Phi^2(z',i\omega_m)}}.
  \label{eq:Elfull2}
\end{eqnarray}
$z=\cos\theta$ and $z'=\cos\theta'$ denote directions on the Fermi
surface, $T$ is the temperature, $g$ is a unitless coupling constant
and $\lambda$ is the unitless interaction function. Note that for
density-density interactions such as we are considering here, the
Eliashberg equations are identical for all singlet and triplet pairing
channels. In the following we shall assume that $\lambda$ takes the
form
\begin{eqnarray}
  \label{eq:CK}
  \lambda(z-z',\Omega) & = & 
  \left[\left((\Omega/\Lambda)^2+(z-z'-2)^2\right)^{K-1}
    \right. \nonumber \\ & & \left. -\left((\Omega/\Lambda)^2+(z-z'+2)^2\right)^{K-1}
    \right] \nonumber \\ & & \frac{\Gamma(1-K)} {(z-z')\Gamma(K)} - \tilde U,
\end{eqnarray}
where $\Lambda=v_fk_f$ is a high energy cutoff, and $v_f=k_f/m$. This
form does not contain the exact Luttinger Liquid correlation function,
but it contains a $2k_f$ power-law divergence, which becomes
logarithmic in the non-interacting case $K\to 1$, and is always
positive. Furthermore, consistency with the $K\to 1$ limit sets
$g=Ak_f^4V^2/16\pi^2\Lambda^2$. The last term, with the unitless
$\tilde U= Uk_f^3/4\pi\Lambda g$, comes from the local repulsion
within the 3D band. By solving the Eliashberg equations,
Eqns. (\ref{eq:Elfull1},\ref{eq:Elfull2}), one obtains the self energy
$\Sigma(z,i\omega_n)$ and pairing field $\Phi(z,i\omega_n)$, which we
assume to be invariant under rotations about the $k_z$ axis, as a
function of the temperature $T$, the effective Luttinger parameter
$K$, and the coupling constants, $g$ and $\tilde U$.

We begin by considering the self energy $\Sigma(z,i\omega_n)$ in the
absence of pairing. Setting $\Phi(z,i\omega_n)=0$, and taking the zero
temperature limit, eq. (\ref{eq:Elfull1}) becomes
\begin{equation}
  \label{eq:Elnormal}
  \Sigma(z,\omega) = g\int_{\omega'}\int_{-1}^1dz'
  \lambda(z-z',\omega-\omega')\,{\rm sign}(\omega').
\end{equation}
In order to determine whether one finds a diverging self energy in
this model, it is convenient to calculate its derivative. The most
diverging term takes the form
$\partial\Sigma/\partial\omega|_{\omega=0} \sim (1\pm z)^{2K-1}$,
which indicates a diverging self energy at $z=\pm 1$ for $K\le 1/2$
(at $K=1/2$ the divergence is logarithmic). Evaluating
$\Sigma(z,\omega)$ itself at these points we find
\begin{equation}
  \label{eq:Spoles}
  \Sigma(z=\pm 1,\omega) \simeq -\frac{2\,g\,\omega}{2K-1}
  \left(2^{2K-1}-\frac{1}{2K}\left|\frac{\omega}{\Lambda}\right|^{2K-1}\right).
\end{equation}
Consequently, we find that the renormalized quasi-particle weight,
$Z(z,\omega)=(1-\partial\Sigma(z,\omega)/\partial\omega)^{-1}$, vanishes at
the ``north and south poles'' of the Fermi surface, which are
connected by the divergent $2k_f$ interaction. However, as far as
pairing is concerned, the main effect of the non-Fermi-liquid behavior
of these ``hot spots'', is to reduce the critical temperature, since,
as we show later, these points are gapped in the superconducting
phase.

Next, we address the Eliashberg equations in the superconducting
phase. It turns out that it is not possible to obtain a triplet
pairing solution, which requires
$\Phi(-z,i\omega_n)=-\Phi(z,i\omega_n)$. To see this, we set $z=1$ in
the RHS of Eq. (\ref{eq:Elfull2}), and note that the main contribution
to the LHS comes from the vicinity of $z=-1$. At these points the two
sides of the equation have the opposite sign, which demonstrates that
there is no triplet solution. Physically, this can be traced back to
the real space oscillations of the correlation function,
$C(\br,\tau)$, with period $2k_f$. Whereas the on site effective
interaction between opposite spin particles, generated by these
fluctuations, is attractive, it is negative at a distance $\pi
k_f^{-1}$, which is the average distance between neighboring particles
of identical spin. Thus, only singlet pairing is expected to emerge
from this effective interaction.  We note, however, that triplet
pairing may arise from multi-orbital interactions when the orbital
content varies as one goes around the Fermi
surface.\cite{wu_triplet_2015,zhou_theory_2015}

Finally, focusing on singlet pairing solutions, we used an iterative
approach to solve for $\Sigma(z,i\omega_n)$ and $\Phi(z,i\omega_n)$ at
finite temperature. To do this we restricted the Matsubara sum to run
only over frequencies below a fixed high energy cutoff,
$-\Lambda<\omega_n<\Lambda$, and divided the $z$-axis into small
segments $\delta z$. Typically, we took $\delta z = 0.04$.

\begin{figure}[t]
  \centering
  \includegraphics[width=\linewidth]{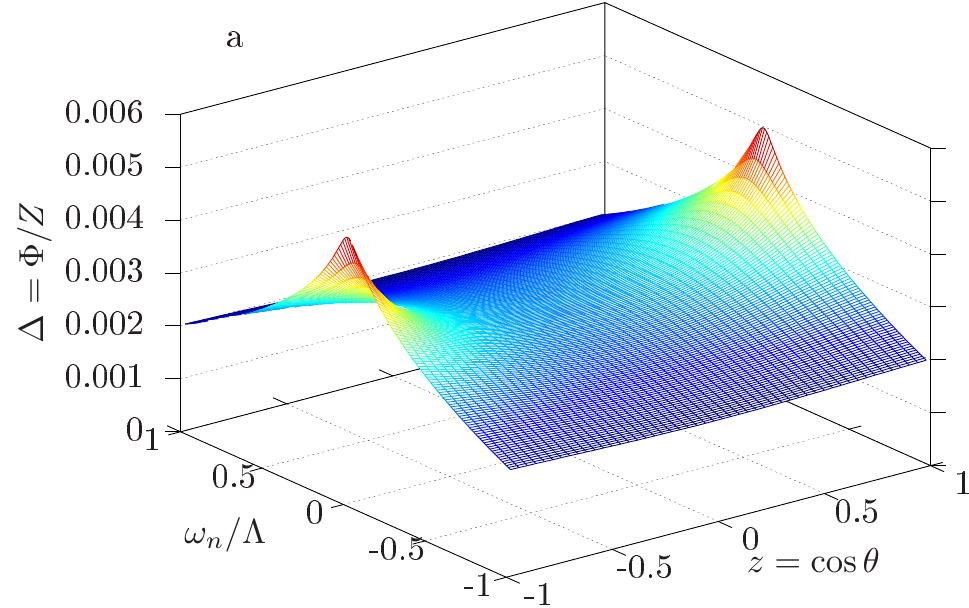} \\
  \includegraphics[width=\linewidth]{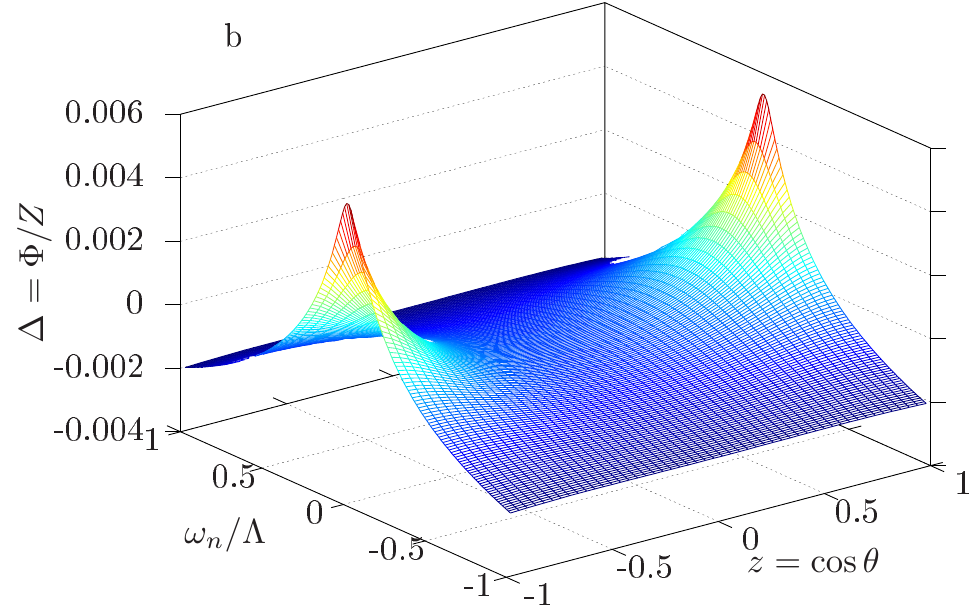}
  \caption{Two examples of $\Delta(z,i\omega_n)$ as obtained by
    numerically solving the Eliasberg equations,
    Eqns. (\ref{eq:Elfull1},\ref{eq:Elfull2}), for $K=0.5$ and $T 
    = 2\times 10^{-4}\Lambda$. (a) $\tilde U = 0$, $g=0.422$. (b)
    $\tilde U = 1$, $g=3.55$.}
  \label{fig:examples}
\end{figure}
\begin{figure*}[t]
  \centering
  \includegraphics[width=\linewidth]{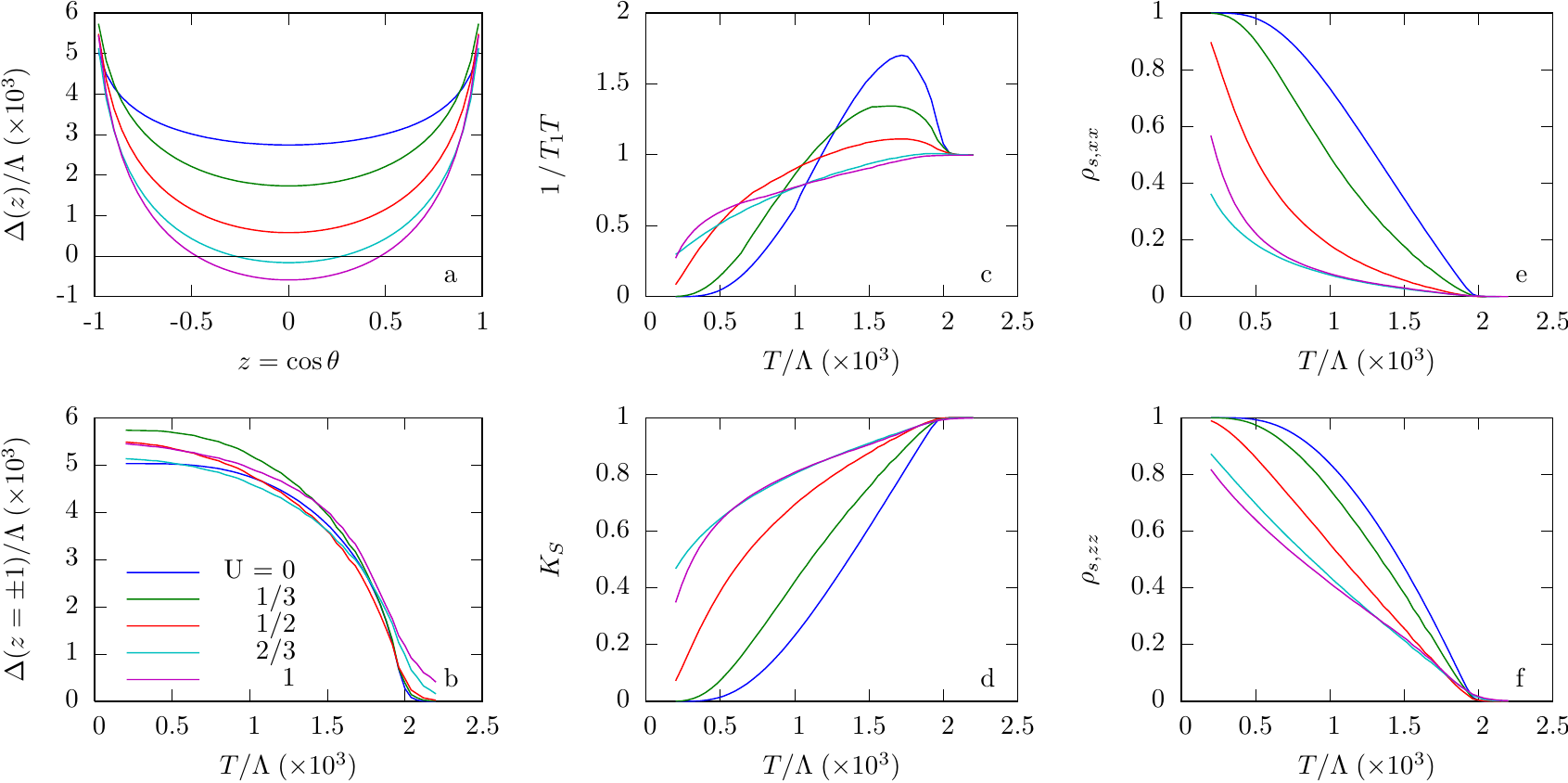}
  \caption{(a) Low $T$ gap function $\Delta(z)$. Nodal lines begin to
    appear for $\tilde U \gtrsim 0.6$. (b) $\Delta(z=\pm 1)$ as a
    function of temperature. $g$ was chosen such that for all choices
    of $\tilde U$, $T_c\simeq 0.002\Lambda$. (c) Nuclear spin
    relaxation rate, $1/T_1T$, normalized to its value at $T_c$.     
    (d) Knight Shift, $K_S$, normalized to its value at $T_c$. 
    (e)-(f) Superfluid stiffness in the $x$ and $z$ directions,
    normalized to their value at $T=0$.}
  \label{fig:A1}
\end{figure*}

As discussed in the introduction, the divergent effective interaction
results in gap functions,
$\Delta(z,i\omega_n)=\Phi(z,i\omega_n)/Z(z,i\omega_n)$ which depend
strongly on the Fermi surface position $z=\cos\theta$, and
unconventional superconducting phases may emerge. Generally, $\Delta$
is peaked near the poles, $z=\pm 1$, and decreases as $z\to 0$, or
with increasing $\omega$. Under some conditions, the pairing function
exhibits nodal lines, defined as the collection of points on the Fermi
surface where $\lim_{\omega\to 0}\Delta(z,\omega)=0$. These occur when
the local repulsion $\tilde U$ is strong enough to suppress the
uniform pairing, $s$-wave-like solutions, and favours
$d_{z^2}$-wave-like solutions, with a pair of nodal rings at $\pm
z_0$. Two example solutions are shown in Figure \ref{fig:examples}.

\section{Experimental consequences}

Unconventional pairing is often manifested in various experimentally
observable quantities. In the following, we show how increasing the
local repulsion $\tilde U$ introduces nodal lines into the gap
function $\Delta$, which in turn dramatically effect several
observables. Specifically, we use the zero frequency limit,
$\Delta(z)\equiv \lim_{\omega\to 0} \Delta(z,\omega)$, as obtained
from our numerical solutions of
Eqns. (\ref{eq:Elfull1},\ref{eq:Elfull2}), to approximately calculate
the real time spin correlation fucntion $\chi_0(\bq, \omega)$ from which
we obtain\cite{bulut_weak-coupling_1992} the Knight shift,
$K_S\sim\lim_{\bq\to 0}\chi_0(\bq,0)$, and spin relaxation time $T_1$,
as given by
\begin{equation}
  \label{eq:T1def}
  \frac{1}{T_1T}\sim \int_\bq\frac{{\rm Im}\chi_0(\bq,\omega)}{\omega}.
\end{equation}
In addition we calculate the superfluid
stiffness\cite{scalapino_insulator_1993} tensor components,
$\rho_{xx}$ and $\rho_{zz}$, which are related to London penetration
depth and $\mu$SR measurements. See Appendix \ref{sec:bdg} for
details. In figure \ref{fig:A1} we show results for several systems
with $K=1/2$ and varying $\tilde U = 0,1/3,1/2,2/3,1$. For each
$\tilde U$ we chose, respectively, $g = 0.422,0.885,1.52,2.4,3.55$
such that the critical temperature $T_c$ remained approximately the
same, i.e., $T_c\approx 0.002\Lambda$, see
fig. \ref{fig:A1}b. Fig. \ref{fig:A1}a depicts the low temperature gap
function $\Delta(z)$, which exhibits nodal lines for $\tilde U\gtrsim
0.6$. When the repulsion in the 1D band is stronger, $K$ decreases and
the nodes appear for lower values of $\tilde U$.  These nodes first
appear at the ``equator'', $z=0$, and move outward towards the
``poles'' as $\tilde U$ is increased. Figure \ref{fig:A1}c shows
$1/T_1T$ as a function of temperature. Fully gapped Fermi surfaces
exhibit a Hebel-Slichter peak below $T_c$, due to constructive
contribution of quasiparticles to the spin relaxation process, as
given by the BCS coherence factors. In our model, the height of the
peak decreases together with the gap minimum, until it vanishes when
the nodal lines are formed at high enough $\tilde U$. In conventional
superconductors there is a drop in the Knight shift below $T_c$.  As
evident from figure \ref{fig:A1}d, this drop becomes less steep for
larger $\tilde U$. Finally, $\rho_{xx}$ and $\rho_{zz}$ decrease
linearly with $T$ at low temperatures when the gap function has nodes,
while they remain roughly constant at low $T$ when the system is fully
gapped. The striking difference bwtween the $T$ dependence of
$\rho_{xx}$ and $\rho_{zz}$, as seen in figures \ref{fig:A1}e,f, is
due to the large anisotropy in the gap function, as one moves on the
Fermi surface from the ``poles'', $z=\pm 1$, to the ``equator'',
$z=0$. Furthermore, a quasiclassical
treatment\cite{volovik_superconductivity_1993}, as well as general
scaling considerations\cite{simon_scaling_1997}, indicate that, in the
presence of nodal lines, the magnetic field dependence of the specific
heat is expected to follow $\sqrt{H}$.  This is attributed to the
linear dependence of the quasiparticle density of states on energy.

\section{Discussion}

The model considered in this paper, given by Eqns. (\ref{eq:S0}) and
(\ref{eq:SI}), is a caricature of the more realistic models studied
previously in the context of the new Cr based superconductors
\cite{xian-xin_magnetism_2015, jiang_electronic_2015,
  zhou_theory_2015, wu_triplet_2015}. It is therefore important to
consider under what conditions our results hold, when the model is
made more realistic. 
In our band structure, Eq. (\ref{eq:S0}), there are two idealized
features, namely, the of transverse dispersion in the 1D band, and the
identical $k_f$ for both 1D and 3D bands. The former ensures the
presence of divergent fluctuations together with the absence of long
range order, while the latter enables the efficient coupling of the 3D
band to the divergent fluctuations.  Relaxing these approximations
cuts-off the $2k_f$ divergence in Eq. (\ref{eq:Seff}), making the
effective interaction more modestly peaked.  Qualitatively, this is
not expected to change the superconducting phase, besides reducing
$T_c$ and increasing the values of $\tilde U$ required to obtain a
nodal ring in the gap function. Other details, such as deviations from
a spherical Fermi surface and differences between the 3D Fermi and 1D
Luttinger velocities, are also not expected to alter the qualitative
picture. Ultimately, however, the addition of weak transverse
dispersion in the 1D band, will make it susceptible to pairing at low
enough temperatures, possibly independent of the pairing in the 3D band.

In the main text we have considered only the effect of density-density
interactions between bands, Eq. (\ref{eq:SI}), which, as discussed in
Appendix \ref{sec:app}, are consistent with the assumption of
Luttinger liquid correlations for the 1D electrons. Such terms are
generic, and exist in any electronic system.  On the other hand,
additional interaction terms may exist, and if dominant enough may
change the nature of the superconducting phase.
Specifically, pair hopping terms are expected, in general, when the
bands have a non trivial orbital content, or, due to a finite Hund's
rule coupling between bands. These interaction terms would induce
pairing in the 1D band by established pairing in the 3D
band. Furthermore, similar conditions may give rise to triplet
superconductivity, as obtained in Refs. \cite{zhou_theory_2015,
  wu_triplet_2015}.  Nevertheless, our results are expected to hold
when these additional terms are weak, and could be neglected.

In summary, we have studied the superconducting phase expected to
arise from $2k_f$ density fluctuations in a model of interacting 1D
and 3D electrons. By solving the Eliashberg equations we have shown
that strongly anisotropic singlet pairing forms on the 3D Fermi
sphere. Nodal rings are formed in the gap functions when strong enough
local repulsion is included.  We have further shown that our simple
model is sufficient to reproduce some of the striking experimental
measurements in the Cr based superconductors, associated with the
existence of nodal rings in the gap function.

\section*{Acknowledgments}
\label{sec:acknowledgments}

We are grateful to Sung-Sik Lee for useful discussion.  This work was
supported by the NSERC of Canada, the Canadian Institute for Advanced
Research, and the Center for Quantum Materials at the University of
Toronto.

\appendix

\section{Feedback of 3D fluctuations on 1D band}
\label{sec:app}

Coupling between 1D and 3D bands affects not only the electrons in the
3D band but also those in the 1D band. It is simpler to study this
effect by visualizing the 1D band as a two-dimensional array of
disconnected chains,
\begin{equation}
  \label{eq:S01D}
  S_0 = \sum_{i\sigma}\int_{k,\omega}\bchi_{i\sigma}(k,\omega)
  \left(-i\omega+\frac{1}{2m} (k^2-k_f^2)\right)\chi_{i\sigma}(k,\omega),
\end{equation}
where $i$ enumerates the chains.  The effective density-density
interaction obtained by integrating out the 3D fermions, to second
order in $V$, is given by
\begin{eqnarray}
  \label{eq:SI1D}
  S_{\rm eff}^1 & \sim & -\frac{V^2}{2}\sum_{ij\sigma\sigma'}\int_{kk'q,\omega\omega'\Omega}
  \bchi_{i\sigma}(k+q,\omega+\Omega)\chi_{i\sigma}(k,\omega)
  \nonumber \\ & & \qquad\qquad   \Pi_{ij}(q,\Omega)
  \bchi_{j\sigma'}(k'-q,\omega-\Omega)\chi_{j\sigma'}(k',\omega').
    \nonumber \\
\end{eqnarray}
Here, $\Pi_{ij}(q,\Omega)$ is defined as
\begin{equation}
  \label{eq:Pi}
  \Pi_{ij}(q,\Omega)\equiv\int dr d\tau\, \Pi(\br_i-\br_j+r\hz,\tau)
  e^{iqr-i\Omega\tau},
\end{equation}
where $\Pi(\br,\tau)$ is the density-density correlation function of
the 3D fermions, and $\br_i,\br_j$ are the two-dimensional coordinates
of chains $i$ and $j$. Within a chain, $i=j$, this effective
interaction modifies the bare forward ($q\simeq 0$) and back($q\simeq
2k_f$)- scattering interaction terms. Nevertheless,
$\Pi_{ii}(q,\omega$) is non-divergent and the 1D correlation functions
are still expected to exhibit the power-law form of
Eq. (\ref{eq:CLL}). On the other hand, the induced density-density
interaction between chains, $-V^2\Pi_{ij}(q,\Omega)$, $i\ne j$, may,
in principle lead to a charge-density ordered state. In this study,
however, we assume that such tendencies are suppressed, for example,
by long-range repulsion terms which we have not included explicitly in
the model. Thus, for our purposes it is reasonable to neglect the
induced coupling between chains.

We finally address the effect of 3D fluctuations on the dispersion of
the 1D electrons. Referring again to the disconnected chains picture,
we note that since the 1D Green's function obeys
$\braket{\bpsi^1_i\psi^1_j}\sim\delta_{ij}$, then also the self energy
must obey $\Sigma^1_{ij}\sim\delta_{ij}$. To see this it is sufficient
to observe that all the diagrams in $\Sigma^1$ of any order contain
only a single open fermion line, connecting the incoming and outgoing
fermions. Consequently, the effective interaction does not induce
dispersion which remains solely along the $k_z$ axis.

\section{Extracting superfluid stiffness and spin correlation function
  from the Bogolubov-de-Gennes spectrum}
\label{sec:bdg}

The task of extracting measurable quantities from the solutions of the
Eliashberg equations, Eqns. (\ref{eq:Elfull1}, \ref{eq:Elfull2}), is
simplified by approximating the frequency dependent gap function by
its zero frequency limit, $\Delta(z)=\lim_{\omega\to
  0}\Delta(z,\omega)$. Beglubov-de-Gennes quasiparticles are
well defined under such an approximation, and it is straight forward to
extract measurable quantities from their spectrum. Thus, the
superfluid stiffness, normalized to its $T=0$ value, is given by
\begin{equation}
  \label{eq:rhos}
  \rho_{ii} \simeq 1 - \frac{3(2\pi)^3\Lambda}{ k_f^5}\int_\bk k_i^2
  \frac{\partial f}{\partial E_\bk}  ,
\end{equation}
while the spin susceptibility is
\begin{eqnarray}
  \label{eq:chis}
  \chi_0(\bq,\omega)\!\! & = &\!\! \frac{1}{4}\sum_{a,b=\pm}\int_\bk
  \frac{f(aE_\bk)-f(bE_{\bk+\bq})}{\omega-aE_\bk+bE_{\bk+\bq}+i0^+}
  \\ \nonumber & &\!\!\! \left\{\left(1+a\frac{\epsilon_\bk}{E_\bk}\right)
    \left(1+b\frac{\epsilon_{\bk+\bq}}{E_{\bk+\bq}}\right)
    +ab\frac{\Delta_\bk\Delta_{\bk+\bq}}{E_\bk E_{\bk+\bq}}\right\}.
\end{eqnarray}
Here, $\Delta_\bk$ is obtained from the solution of the Eliashberg
equations, $\epsilon_\bk=k^2/2m$,
$E_\bk=\sqrt{\epsilon_\bk^2+\Delta_\bk^2}$, and $f(E_\bk)$ is the
Fermi function. Note the BCS coherence factors which enter $\chi_0$
and which are responsible for the appearance of the Hebel-Slichter
peak below $T_c$ in fully gapped superconductors.

\bibliography{1d3d}

\end{document}